\documentstyle[11pt,newpasp,twoside,epsf]{article}
       	\newcommand{\citeauthor}[1]{\def\citeauthoryear##1##2##3{\rm ##1}\cite{#1}}
     	\newcommand{\citeyear}[1]{\def\citeauthoryear##1##2##3{\rm ##3}\cite{#1}}
        \newcommand{\citet}[1]{\citeauthor{#1} (\citeyear{#1})}
        \newcommand{\citealt}[1]{\citeauthor{#1} \citeyear{#1}}

\def\edcomment#1{\iffalse\marginpar{\raggedright\sl#1\/}\else\relax\fi}
\reversemarginpar

\begin{document}
\title{The magnetism of the very quiet Sun}
 \author{J. S\'anchez Almeida}
\affil{Instituto de Astrof\'\i sica de Canarias, 
	La Laguna, Tenerife, Spain}
\begin{abstract}
	When the polarimetric sensitivity and the angular
	resolution exceed a threshold, magnetic fields show up
	almost everywhere on the solar surface. 
	Here I revise the observational properties of 
	the weakest polarization signals,
	which appear in the InterNetwork (IN) regions. 
	We already have some information
	on the magnetic field strengths and inclinations, 
	mass motions, lifetimes, magnetic fluxes, magnetic energies, etc.
	Since the IN covers a substantial faction
	of the solar surface, it
	may account for most of the unsigned magnetic flux 
	and energy existing on the solar surface at any given time.
	This fact makes IN fields potentially important to understand the 
	global magnetic properties of the Sun
	(e.g., the structure of the quiet
	solar corona, an issue briefly addressed here).
	The spectropolarimeters on board of Solar-B have the resolution
	and sensitivity to routinely detect these IN fields.
\end{abstract}

\section{Introduction}

	Most of the solar surface appears non-magnetic when it is 
	observed in routine synoptic magnetograms. However, magnetic 
	fields are detected almost everywhere, also in InterNetwork (IN) 
	regions, when the polarimetric sensitivity and the angular
	resolution are high enough. These magnetic fields are 
	now accessible to many spectropolarimeters.
	They cover much of the solar surface and, therefore,
	they may account for most of the unsigned magnetic flux 
	and energy existing on the solar surface at any given time.
	The contribution summarizes the main observational properties
	of the IN fields, as deduced from these recent measurements. 
	In addition, it discusses the origin of the IN magnetism,
	and why it is a natural target for Solar-B.

\section{Operative Definition of Quiet Sun Magnetic Fields}
\begin{figure}
\plotfiddle{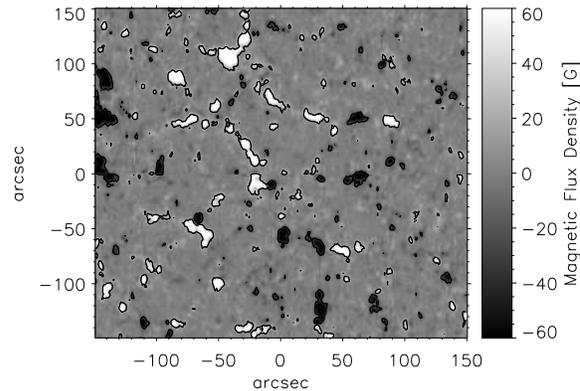}{4.0cm}{0}{45}{45}{-160}{-180}
\caption{Routine Kitt Peak magnetogram of the disk center taken
	during the solar minimum. Most signals correspond to the network,
	whose 
	features have been outlined with a contour (at 20~G). 
	Pixels outside the contour occupy 92\% of the surface.
	}
\label{quiet_mag}
\end{figure}

	There is no universal consensus on what the term 
{\it quiet Sun magnetic fields} really means. Here it will
denote those regions which 
do not show significant polarization signals in the traditional
synoptic magnetograms (e.g., the gray background in
Fig. \ref{quiet_mag}). These regions are often called 
InterNetwork or IntraNetwork (IN). The separation between network and IN
is not clear. It depends on the polarimetric
sensitivity of the  measurement. When the sensitivity is high enough,
most of the surface produces
polarization and the network and the IN become indistinguishable.
Since the observation of the IN requires this high sensitivity,
there is always some degree of ambiguity in the separation
(there is some contamination of the IN signals with network signals).
Keep in mind this caveat\footnote{Eventually,
the standard magnetograms will routinely detect 
IN fields in full. Then the somehow artificial
separation between network and IN will be abandoned,
and we will talk about {\it quiet Sun magnetic fields} with
{\it high flux density} (network) or {\it low flux density} (IN).
}.

\section{Surface Coverage\label{scoverage}}

\begin{figure}
\plotfiddle{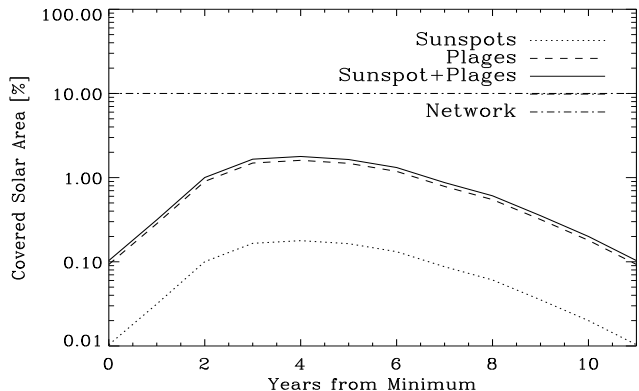}{4.5cm}{0}{50}{45}{-160}{-175}
\caption{Variation of the solar surface covered by sunspots
and plage regions along the cycle. Note that even
during the solar maximum, most of the traditional 
full disk magnetograms appear devoid of
magnetic structures.}
\label{coverage}
\end{figure}

The quiet Sun IN occupies most of the solar surface, even
during solar maximum. Figure \ref{coverage}
shows the area covered by sunspots and plage regions
as a function of time\footnote{
It has been adapted from
\citet{har00}, using the mean Sunspot number plus
the scaling between sunspot number and area 
provided by a typical sunspot group. It has
a sunspot number equals to 12, and it covers
200 millionths of solar hemisphere with sunspots
and 1800 millionths with plages.}. 
Figure \ref{coverage} also includes a 10\% level to indicate
the fraction of surface corresponding to the
network (e.g., 
the contours in Fig. \ref{quiet_mag} outline 
some 8\% of the area in this Kitt Peak magnetogram).
The level is shown as a constant since
network signals in conventional
magnetograms 
do not seem to suffer strong variations
with the cycle (see, e.g., \citealt{har93} , Chapter 12, Fig. 9). 
Figure \ref{coverage} reveals 
that some 90\% of the solar surface is IN quiet Sun,
even during the maximum. 

\section{Complex Topology of the Quiet Sun Magnetic Fields\label{topology}}
The spatial resolution of the present or even forthcoming observations
is not enough to resolve the structure of the
quiet Sun magnetic fields. In other words,
the physical properties of the field (e.g., strength or
direction) vary  
within resolution elements which seldom reach
0\farcs 5. 
These variations
are so large that often the mean value of a physical
quantity bears no information on the actual value of the quantity.
Many different observations prove the extreme
disorder existing in each resolution element.

The polarization of the spectral lines always shows asymmetries
(\citealt{san96};
\citealt{sig99};
\citealt{san00};
\citealt{kho02}). In particular, the spectral
lines generate net circular polarization, which
would be impossible if either the magnetic field or the 
velocity were uniform. Moreover, 
the variations of magnetic field and velocity
have to occur along the line-of-sight
(e.g., \citealt{san98c}). Since the photospheric lines
are formed in, say, 100 km or 150 km, large
gradients must take place  within such small distances.

\begin{figure}
\plotfiddle{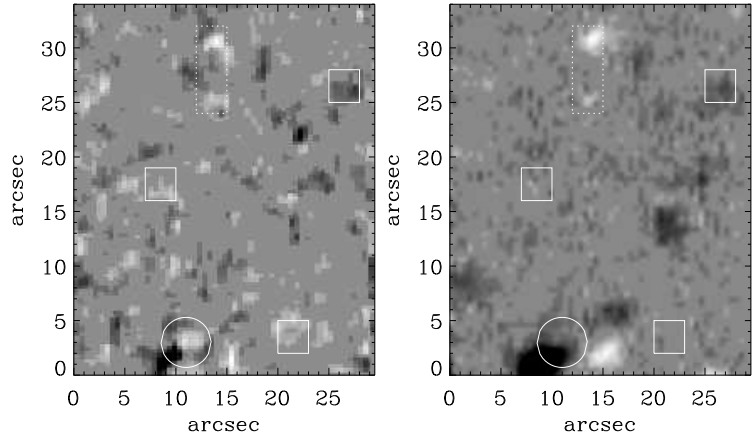}{5cm}{0}{50}{50}{-160}{-25}
\caption{Simultaneous and co-spatial IR (left) and  
visible (right) magnetograms of a quiet Sun IN region
at the disk center.
Often the polarity of the two magnetograms is opposite (e.g., 
see the region within the circle). 
}
\label{irvis}
\end{figure}

The magnetometry of the quiet Sun based on visible 
lines 
is seemingly incompatible with the InfraRed (IR)
magnetometry. For example, Fig. \ref{irvis} 
shows two co-spatial (1\farcs5) and simultaneous (1 min) 
magnetograms obtained with Fe~{\sc i}~6302.5~\AA\ and
Fe~{\sc i}~15648~\AA\ (\citealt{san03c}). Often the
polarity of the visible and IR signals is opposite (see, e.g.,
the circle) indicating that the {\it direction} of the magnetic
field is not defined.  The true field 
has many different orientations
in the resolution element. 
The existence of unresolved opposite polarities 
is well documented in the literature. They
are needed to explain  the asymmetries mentioned
above (\citealt{san00}), and  
the bias of magnetic field inclinations observed away from
the solar disk center (\citealt{lit02}). They also emerge naturally
in the numerical simulations of magneto-convection
(e.g., \citealt{cat99a}; \citealt{ste02}).

The amount of magnetic flux in quiet Sun features
increases exponentially
with increasing spatial resolution. It goes from 1~G for
2--3\arcsec to some 20~G at 0\farcs5 
(e.g., \citealt{san03}, Fig. 3;
\citealt{dom03b}, Fig. 12). These systematic changes would not happen unless
much IN structure remains unresolved.

The Hanle depolarization signals of photospheric lines 
are consistent with a turbulent magnetic field
occupying most of the volume. Although the field strength
is a model dependent quantity, the estimates
require fields certainly different from zero and
smaller than some
130~G (cf., \citealt{fau95}; \citealt{bia99};
\citealt{shc03}; \citealt{tru04}).

Two main consequences arise from the combination of complex
magnetic field topology and limited spatial resolution.
First,
the measurements are bound to underestimate the magnetic
flux and  energy existing in the quiet Sun.
The polarization signals tend to cancel when the
structure is not resolved\footnote{Not true for the 
Hanle signals; see, e.g., \citet{ste94}.}.
Second,
the measurement of the magnetic field properties turns
out to be non-trivial. Extracting physical information
from the observed polarization
involves modeling and assumptions
on the underlying atmosphere. The measurements become
model dependent and non-unique. This is an unavoidable
tribute that must be paid.

\section{Magnetic Field Strength\label{mfs}}
	The traditional solar magnetic structures have  
magnetic field strengths larger than 1~kG. 
It varies from 2.5~kG to 1~kG when
going from the sunspot umbra to the penumbra (e.g., \citealt{bra64}). 
Magnetic concentrations of plage and network regions have 
a field strength between 1~kG and 2~kG (e.g., \citealt{sol93}).
The situation is very different in the quiet Sun, where 
magnetic field strengths spanning more than three orders of magnitude
have been detected. Observations show  field strengths going all the way from
zero to 2~kG.
Those measurements based on visible
lines tend to show kG (e.g., \citealt{sig99}; \citealt{san00}; \citealt{soc02}).
Infrared line based measurements prefer hundreds of G (e.g.,
\citealt{lin99}; \citealt{kho02}). Finally, 
the observed Hanle depolarization
signals demand even weaker fields, of the order of tens of G (e.g.,
	see the references in \S~\ref{topology}).
All these different values are probably consistent with a single
distribution having all field strengths between 0~G and 2~kG.
Different observations, using different
spectral ranges and physical principles, are only sensitive 
to part of the true distribution. Physical mechanisms responsible
of the selectivity of the measurements have been put forward. Hanle 
signals cannot sense Zeeman splittings larger than the natural
width of the atomic levels, which corresponds to some 100~G for the 
transitions  employed in photospheric measurements (e.g., \citealt{fau95}).
On the other hand,  the
large Zeeman splitting of the IR lines smears
the polarization signals arising from kG fields which, however,
can
be observed with lines of smaller splitting, i.e., with visible lines 
(\citealt{san00}; \citealt{soc03}).
\begin{figure}
\plotfiddle{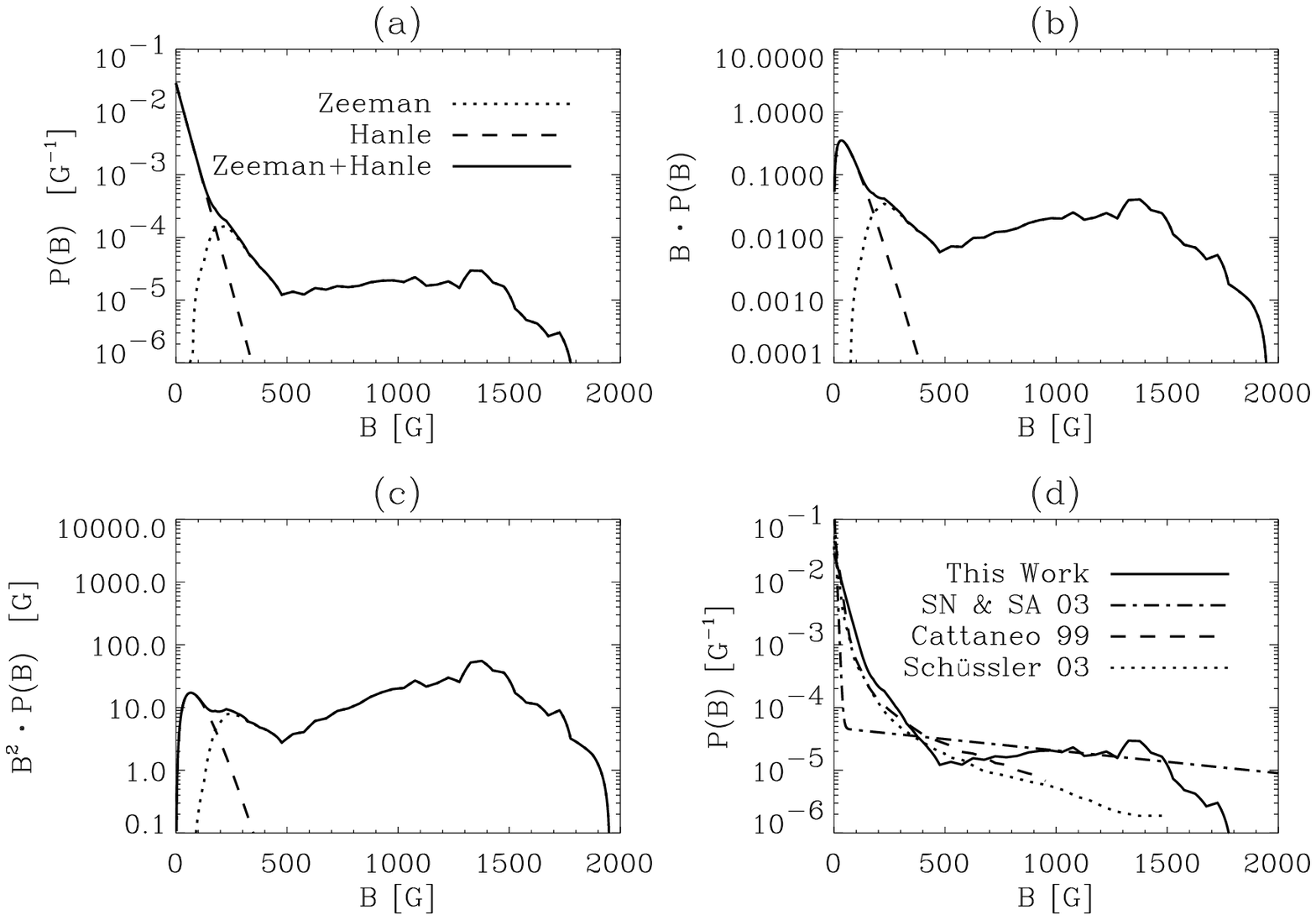}{10cm}{0}{70}{80}{-215}{-300}
\caption{(a) Probability density function ($P$) of finding  
a magnetic field strength $B$ in the quiet Sun (the solid line).
This probability includes a contribution from measurements
based on the Zeeman effect (the dotted line), and another 
arising from Hanle signals (the dashed line). (b) $B \times P(B)$, or
the unsigned flux density per unit of field
strength. 
The contributions from Zeeman and Hanle signals are coded
as in (a).
(c) $B^2 \times P(B)$, or
the magnetic energy density per unit of field
strength. 
The contribution from Zeeman and Hanle signals are coded
as in (a).
(d) $P(B)$ in (a), together with some other PDFs existing
in the literature. 
	(SN\&SA $\equiv$ Socas-Navarro \& S\'anchez Almeida 2003.)
}
\plotfiddle{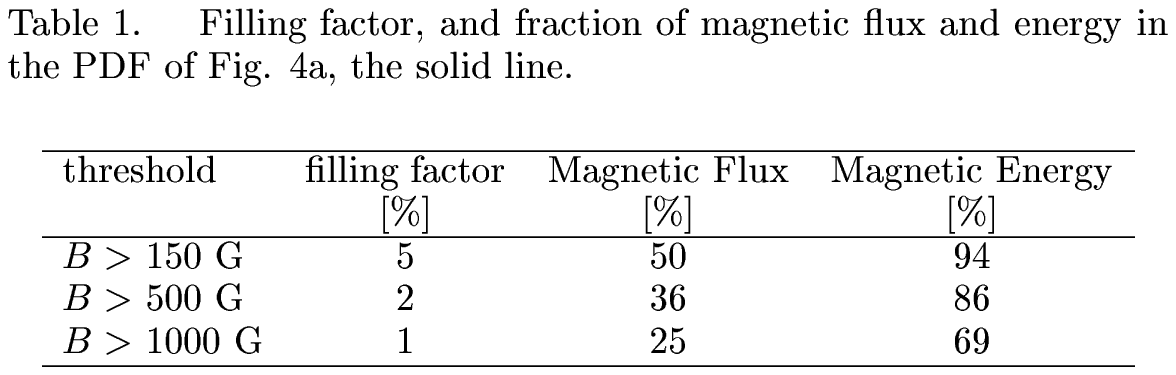}{9cm}{0}{100}{100}{-300}{-480}
\label{pdf}
\end{figure}
The field strength of the quiet Sun needs to be 
characterized with a Probability Density Function (PDF). 
This function,  $P(B)$, 
gives the probability of finding a field strength $B$. 
Although we still lack of a reliable PDF, 
tentative PDFs based on either Zeeman or Hanle signals
have been suggested. As explained above,
Hanle and Zeeman signals only
provide partial information on the true PDF. 
For illustrative purposes,
I have constructed a PDF 
that tries to accommodate both the Zeeman and the Hanle signals.
It is shown in Fig. \ref{pdf}a (the solid line).
This $P(B)$ has been
set up using the PDF from visible and IR Zeeman signals in \citet{san03c},
$f_Z(B)$, plus an exponential\footnote{The
exponential shape of the Hanle PDF is suggested by the
weak fields in 
numerical simulations of magneto-convection (e.g. \citealt{cat99a}), and 
it has been used to model Hanle signals (\citealt{shc03}; \citealt{tru04}).}
component to account
for the Hanle signals,
\begin{equation}
P(B)=w B_0^{-1}\exp(-B/B_0) + (1-w) f_Z(B),
\end{equation}
with $f_Z$ normalized to unity.
The two unknowns of $P(B)$, $w$ and $B_0$, can be set 
from the two first moments of the distribution,
\begin{eqnarray}
<B>=&\int_0^\infty B\cdot P(B)dB=w B_0 + (1-w) \int_0^\infty B\cdot f_Z(B)dB,\cr
<B^2>=&\int_0^\infty B^2\cdot P(B)dB=2 w B_0^2 + 
	(1-w) \int_0^\infty B^2\cdot f_Z(B)dB.
	\label{thiseq}
\end{eqnarray}
$<B>$ and $<B^2>$ have a clear physical interpretation
and can estimated from observations.  
The mean, $<B>$, is related to the unsigned flux density of the quiet Sun 
as measured with magnetograms\footnote{For example,
assuming that the three components of the magnetic
field are identical, then
$<B>$ is $\sqrt{3}$ times the
unsigned flux density measured with longitudinal magnetograms.}.
On the other hand, $<B^2>$ is proportional to the mean
magnetic energy density
in the quiet Sun. Figure \ref{pdf}a has been produced using
\begin{equation}
<B>={\rm 60~G},~~~~~
<B^2>^{1/2}={\rm 170~G}.
	\label{values}
\end{equation}
The value for $<B>$ represents an educated guess based on the average
fields deduced from 
Hanle signals (e.g., \citealt{shc03}), as well as on the true 
unsigned flux
in numerical simulations whose synthetic polarization is
compatible with observations (\citealt{san03}).
The value for the second moment corresponds to a magnetic energy
equals to
a sizeable fraction
of the solar granulation kinetic energy,
explicitly,
\begin{equation}
<B^2>/(8\pi)=\alpha <\rho U^2>/2,
\end{equation}
with the fraction $\alpha = $ 20\%\footnote{It
would be difficult to exceed
this 20\%, which is the level of magnetic energy generated by the 
highly efficient turbulent dynamo of \citet{cat99a}.},
the density $\rho = 3\times 10^{-7}$~g~cm$^{-3}$,
and the velocity $U = $2~km~s$^{-1}$. 
According to the PDF in Fig. \ref{pdf}a,
most of the plasma has very weak magnetic fields. 
Some 98\% of the quiet Sun has a field 
weaker than 0.5~kG (see Table 1). However, the 2\% tail of kG field 
is quantitatively important since it carries a significant
part of the magnetic flux and energy. 
See \S~\ref{energy}
	
	So far no distinction has been made 
between the magnetic field strength in granules and
intergranules. The possible differences have to be confirmed
in the future, however,
kG fields seem to 
prefer intergranules  whereas the weak fields are
associated with granules (see, \citealt{soc04}; \citealt{dom03a}).
This behavior is actually predicted by the numerical simulations
of magneto-convection.
	
\section{Magnetic Flux Density and Magnetic Energy Density\label{energy}}

As the previous section points out, the two first moments
of the PDF
are closely related to the unsigned flux density
measured with longitudinal magnetographs (1st moment),
and with the magnetic energy density (2nd moment). 
Due to the qualitative 
character of the forthcoming discussions, 
$<B>$ is identified  with the unsigned flux density,  and 
$<B^2>/(8\pi)$ with the energy density. Figures \ref{pdf}b and
\ref{pdf}c contain the PDF multiplied by $B$ and $B^2$, respectively.
They represent the unsigned magnetic flux and the magnetic flux density
corresponding to each field strength. The multiplication by $B$ or
$B^2$ suppresses the peak at $B=0$ and, consequently,
despite the fact that most of the quiet Sun has $B\sim 0$~G,
these weak fields do not contribute to the flux and energy.
In other words, the feeble tail of strong fields 
is not negligible but, rather, it determines the magnetic properties
of the quiet Sun.
To be more quantitative, only 5\% of the surface has field strengths
larger than 150~G. However, this 5\% carries 50\% of the flux and
holds 94\% of the magnetic energy (Table 1).
Obviously, these values will be modified upon improvement
of the observational PDF. Nevertheless, significant amounts of strong fields 
are already inferred from the Zeeman magnetometry 
(e.g., \citealt{dom03a}), implying that the tail
of strong fields is bound to play a major role.

	The total magnetic flux density and energy 
of the PDF in Fig. \ref{pdf}a have been imposed artificially.
However, the chosen values are  realistic since
they are based
on reasonable extrapolations of observed quantities. If they were
real, the quiet Sun magnetism would be really important. 
A quiet Sun having as 
much as $<B>$=60~G
carries 5 times more flux than all
active regions at solar maximum, mainly 
because it covers most of the solar
surface (\S~\ref{scoverage}). 
Refer to \citet{san02b} for details on this issue.

\section{Motions and Lifetimes\label{motions}}

The strong kG fields are associated with intergranules
(\S~\ref{mfs}) and, consequently, with downflows.
However, the downflows primarily occur
outside the magnetic concentrations (see \citealt{san00}). 
The motions of the weaker fields are more uncertain,
but they are probably related to 
granules and upflows (see \citealt{soc04}).
The strong IN fields are dragged by horizontal
plasma motions (e.g., \citealt{zha98}; \citealt{dom03b}). 
This is 
the reason why the network (i.e., the supergranulation)
shows in the magnetograms. Actually all spatial
scales of organized photospheric motions appear in the quiet Sun
magnetograms, including the granulation and  the
mesogranulation (\citealt{dom03b}; \citealt{dom03c}).
In particular, the mesogranulation is fairly easy to
detect, whereas it remains elusive when using conventional techniques.

The quiet Sun magnetograms are observed
to evolve on a timescale similar to that of the
granulation (say, 10~min; see, e.g., \citealt{lin99}). The mesogranular
pattern lasts at least half an hour (\citealt{dom03b}).
Moreover, the lifetimes of the large IN patches 
is of the order of a few hours (\citealt{zha98b}). 
Two comments are in order. In spite of the uncertainties
of the lifetimes,
it is very clear that the IN timescales are much
shorter than those characterizing the evolution of active
regions (several days; see, e.g., \citealt{har93}).
Second, the persistence of a signal in a magnetogram does not 
necessarily imply the survival of a single magnetic structure
during this period. It may well be that the flow pattern 
dragging magnetic features persists this long 
(e.g., \citealt{ras03}).
	
\section{Variations with the Solar Cycle\label{cycle}}

	Due to the
potential importance of the quiet Sun magnetism (\S~\ref{energy}), 
its variations along the cycle are of particular interest. 
Little is known about this issue, though.
\citet{fau01} find a factor 2 variation of the mean field.
\citet{san03d} claims no variation within his error bars 
(40\%). \citet{shc03} also find no variation.
Based on this limited information, we can conjecture
that the quiet Sun magnetic flux does not seem to suffer large
variations along the cycle. If it changes, the variations are
far smaller than those observed in active regions, whose 
total flux varies by more than one order of magnitude (e.g.,
\citealt{har93}, Chapter 12, Fig. 4).

\section{Origin of the Quiet Sun Magnetism\label{origin}}
Several possibilities have been 
put forward. The IN may result from the
decay of active regions. This possibility has problems
due to the large amount of magnetic flux and the
short decay time of the IN fields as compared to the
active regions (see \citealt{san03b}
for a quantitative argumentation). The quiet Sun may 
be generated by a surface turbulent dynamo which  transforms
part of the convective kinetic energy into magnetic energy
(\citealt{pet93}; \citealt{cat99a}). This local dynamo
has been criticized by \citet{ste02}, arguing that the
mass flows of the granulation are not restricted to a narrow
layer close to the solar surface but involve the whole
solar convection zone. These authors propose that the IN fields
are produced by the turbulent component of the global solar
dynamo. The numerical simulations inspired in the turbulent
dynamo scenario (no matter whether it is local or global)
produce a complex  magnetic field resembling
the observed IN fields (see, e.g., \citealt{san03}).

\section{Conclusions and Final Comments}

	The quiet Sun is the component of the 
solar surface magnetism that seems to account for most of 
the magnetic flux and magnetic energy (\S~\ref{energy}).
This fact makes it potentially important
to understand the global magnetic properties of the Sun
(solar dynamo, coronal heating, origin of the solar wind, and
so on).
However, its influence has been neglected so far.
It produces very weak polarization signals which 
hardly show up in conventional magnetograms.
The situation is slowly turning around,  and it
will dramatically change with the advent of space-borne polarimeters
like those of Solar-B. For example, Fe~{\sc i}~6302~\AA\ is expected
to produce a circular polarization of the order
of $5\times 10^{-3}$ when the angular resolution reaches 0\farcs 5
(\citealt{dom03b}). These signals are well above the noise level
in the normal mapping model of the SOT spectropolarimeter (\citealt{shi04}).
In other words, Solar-B is expected to routinely detect the quiet Sun magnetic
fields (at least that fraction having kG field strengths, 
see \S~\ref{mfs}).

The true role of the quiet Sun magnetism
is still unknown. Only preliminary steps 
to figure it out have been given.
Let me point out a recent 
work by \citet{sch03b} where they study the influence of the
quiet Sun magnetic fields on the extrapolation of the 
photospheric field to the corona. They conclude that 
an important modification of the network-rooted
field lines is induced by the presence of the IN, implying that
a significant part of this disorganized IN photospheric
field does indeed reaches the quiet corona. 

\acknowledgements
Thanks are due to the Solar B 5  SOC,
and to various colleagues;
F. Kneer, H. Socas Navarro
and 
I. Dom\'\i nguez Cerde\~na, who also set up Fig. \ref{irvis}.
The work has been supported by grants AYA2001-1646,
and EC HPRN-CT-2002-00313.
The ISO/Kitt Peak data in Fig. \ref{quiet_mag} are produced  by
NSF/NOAA, NASA/GSFC, and NOAA/SEL.

\def\nat{Nature}
%

\end{document}